\documentclass[aps,prl,twocolumn]{revtex4-1}
\usepackage{amssymb}
\usepackage{amsmath,bm}
\usepackage{graphicx,color}
 \usepackage{epstopdf}
\usepackage{epsfig}
\usepackage{undertilde}
\usepackage{multirow}
\usepackage[cal=pxtx]{mathalfa}
\usepackage{accents}
\DeclareMathAccent{\wtilde}{\mathord}{largesymbols}{"65}

\usepackage{stackengine}
\stackMath
\newcommand\tenq[2][1]{%
	\def\useanchorwidth{T}%
	\ifnum#1>1%
	\stackunder[0pt]{\tenq[\numexpr#1-1\relax]{#2}}{\scriptscriptstyle\sim}%
	\else%
	\stackunder[1pt]{#2}{\scriptscriptstyle\sim}%
	\fi%
}

\renewcommand{\vec}[1]{{\underline{#1}}}
\newcommand{\B}[1]{{\underline{#1}}} % XXX this is useful to make complex tensor homogeneous XXX

\def\uvec#1{{\hat{\underline{#1}}}}
\def\utilde#1{\underaccent{\wtilde}{#1}}
\def\uttilde#1{\underaccent{\wtilde}{\underaccent{\wtilde}{#1}}}
\begin{document}

\title{Spatial stress correlations in strong colloidal gel}
% \title{Frictional Granular Matter: Protocol Dependence of Mechanical Properties}
%\thanks{Corresponding authors: saikat.roy@iitrpr.ac.in, \\\\\\\\\ chandanamondal2011@gmail.com\\}%
\author{Divas Singh Dagur$^1$,Chandana Mondal$^{2,*}$, Saikat Roy$^{1,}$}
\email{Corresponding authors: saikat.roy@iitrpr.ac.in,\\\,\,\,\,chandanamondal2011@gmail.com}
\affiliation{$^1$ Department of Chemical Engineering, IIT Ropar, Rupnagar, Punjab, India 140001\\$^2$ UGC-DAE Consortium for Scientific Research, University Campus, Khandwa Road, Indore 452017, India.  }

\begin{abstract}
 In this work, we systematically investigate for the first time the nature of stress correlations in soft colloidal gel materials which support tensile and compressive forces as well as finite rolling torque, as a function of system pressure. Similar to previous studies on frictional granular matter with only compressive forces and without any rolling torque, the full stress autocorrelation matrix is dictated by the pressure and torque autocorrelations due to mechanical balance and material isotropy constraints. Surprisingly, it is observed that the gel materials do not behave as a normal elastic solid {\it close to the gel point} as assumed loosely in the literature because the real space pressure fluctuations decay slower than the normal. We also demonstrate that at low pressure the fractal like structural correlation determines the pressure fluctuations and this is manifested in the real space in terms of inhomogeneous and anisotropic force networks formed due to large voids. Far away from the gel point, as the voids collapse under compression, the force chain network becomes homogeneous and isotropic and the pressure fluctuations become normal leading to normal elastic decay at long range, behaving similar to frictionless granular matter and glass. We also observe that the torque autocorrelation is not hyperuniform in the presence of rolling resistance close to the gel point. Furthermore, we link the abnormal pressure fluctuations to the non-hyperuniform behaviour of the system with respect to the local packing fraction fluctuations, thus relating the deviations from the normal elastic behaviour across various non-equilibrium systems under a common framework.

\end{abstract}
%Although the normal/anomalous elasticity is attributed to the normal/slower than normal elastic decay, there is no common microscopic explanation for the same.
%Amorphous materials with the central interaction show elastic like long range stress correlation as observed in supercooled liquids, colloids and frictionless granular materials. The deviation from the normal elastic behavior is recently observed in the case of dry frictional granular materials which support only compressive forces.

\maketitle
Colloids form disordered solids in seemingly different ways based on the inter-particle potential and the volume fraction. At high volume fraction, crowded interaction between hard sphere colloids gives rise to caging effect and dynamical arrest which results in repulsive colloidal glass. Increasing the strength of attraction leads to the formation of attractive glass. In contrast, at  very low volume fraction and strong attraction, colloids form space filling percolated networks consisting of fractal clusters and manifest solid like properties. One of the challenges in the soft matter science is to unravel the physics of amorphous solids such as glass, gels, granular materials under a unified framework.\\
Amorphous solids unlike the perfect crystals have many microscopic degrees of freedom in their equilibrium states as it is not necessarily the global minimum of the potential energy surface rather it is at some local minima. These metastable states are known as inherent states and characterization of these states are difficult as the system properties are strongly dependent on the preparation protocol. Although properties of the inherent states vary across different amorphous media, rare universalities in some of the macroscopic properties are found. Simulations on model glass\cite{wu2017anomalous}, granular matter\cite{henkes2009statistical,wu2017anomalous} and supercooled liquids\cite{lemaitre2014structural,wu2015anisotropic,lemaitre2015tensorial} show a power law decay of  spatial shear stress correlation ($1/r^d$ in $d$ dimension) with quadrupolar anisotropy. 
%Experimental investigations of strain correlations in colloidal glass \cite{chikkadi2011long,jensen2014local,illing2016strain} 
%and granular matter\cite{le2014emergence} also display similar long range correlations.
% One of the plausible explanation for the development of such long range correlations was that the elastic relaxation of the so called Eshelby transformations\cite{lemaitre2014structural}. Similar stress correlations are also directly measured in experiments with photoelastic granular discs \cite{sarkar2013origin,bi2013fluctuations} where the elastic range is infinitesimally small. These observations raised doubts over the explanation based on the Eshelby transformations which is elastic in nature. 
%On the theoretical front, \cite{lemaitre2017inherent} Lemaitre showed that the mechanical balance and the material isotropy are sufficient to capture these long range anisotropic stress correlations without the involvement of any elasticity for the Hamiltonian systems.
% \citet{henkes2009statistical} used a field theoretic model to describe the spatial stress correlations in granular matter. They showed that the pressure fluctuations in Fourier space approach a constant in the small wave vector limit and the simple shear stress shows $1/r^2$ decay in space, which was in line with the numerical simulations\cite{lois2009stress,wu2015statistics} and experiments\cite{lois2009stress} on granular disks. \citet{degiuli2018field,degiuli2018edwards} constructed a field theory to predict the long range correlations based on the constraint of mechanical equilibrium alone.
On the theoretical front, Lemaitre showed that mechanical balance and material isotropy \cite{lemaitre2021frictional,lemaitre2021stress,lemaitre2017inherent}
constrain the entire stress correlation matrix to be fully determined by two spatially isotropic functions, pressure and torque auto-correlations in frictional granular materials. Another interesting observation was that for the frictionless granular matter, the pressure autocorrelations follow normal elastic decay($1/r^2$) at long distance whereas a divergence is seen in the frictional case due to slower than normal decay of the pressure fluctuations. \\
%Slower than normal pressure fluctuations can be explained by the quenched correlations developed during compression of frictional matter well below the jamming transition.
Although simulation data on granular matter and model glass support the theoretical predictions, no study has been devoted to understand the nature of stress correlations and applicability of the existing theories in strongly attractive systems such as colloidal gels for which thermal fluctuations are negligible compared to the strong attractive potential, $-U/k_BT\geq20$. Colloidal gel forms, at low volume fraction ($1\%$), a stable percolating network which is isotropic and at mechanical equilibrium\cite{roy2016yielding}. The network is fractal at low volume fraction and becomes attractive glass at high packing fractions\cite{joshi2014dynamics} and consequently looses fractal correlations. Due to its unique mechanical and transport properties, gels find applications in diverse areas like foods, pharmaceuticals, cosmetics, aerogels and drug delivery. Hence, it will be of great interest to find out whether these systems present long range correlations or not since it has been recently found that local instabilities generated during mechanical perturbations
induce large scale plastic events via long range stress correlations in colloidal glass\cite{chikkadi2011long}. It is also important to test whether the present theory \cite{lemaitre2021frictional,lemaitre2021stress} can capture the stress correlations in this type of soft gel materials.
In strongly aggregated colloidal gel, the formation of a stable percolating network at the gel point requires the structural organization of the network in such a way so that the mechanical balance is satisfied both at the particle and macroscopic level. The fundamental physics behind the emergence of elasticity and its nature close to the gel point is poorly understood. It begs for a detailed investigation into the microscopic nature of stress fluctuations in the colloidal gel network to find out whether the stringent requirement of mechanical balance imposes some long range stress correlations at the gel point and  whether the nature of this correlations changes at high volume fractions.\\ In this letter, we show via numerical simulations that the stress correlations in colloidal gel is fully determined by two spatially isotropic functions: namely the pressure and torque auto-correlations. We find that pressure auto-correlations show divergence at small wave number close to the gel point and far away from it, the slope of the divergence decreases. It implies that the elastic theory which is generally applied to the gel network at low pressure must be revisited.  
%The slope of the divergence in pressure auto-correlations is in line with the measured pressure fluctuations as a function of the system's size.
Although the gel network is formed at very low packing fractions compared to granular materials, the symmetry and nature of the stress autocorrelation matrix show remarkable universal features. \\
\textit{Methodology and theoretical background}: The simulation starts with the random placement of soft elastic discs of diameter $D$ in a \textit{two-dimenional} periodic box at very dilute packing fractions. It was shown recently \cite{roy2016universality,roy2016yielding,roy2016b}that the essential physics of the colloidal gel can be captured by modeling them as cohesive granular matter. We employ `Discrete element method'\cite{cundall1979discrete} to simulate the  system via open source codes\cite{plimpton1995fast}. The normal contact deformation is modeled as Hookean with a constant attraction force acting centrally when a contact is made. Both the sliding and the rolling resistance between the particles \cite{pantina2005elasticity,furst2007yielding} are considered to capture the behavior of real colloidal gel. For the details of the contact interactions, the reader is referred to Ref. \cite{roy2016yielding}. The particles are given random kicks to its translational and rotational degrees of freedom compensated by a damping term. The Langevin equation is solved to update the particle positions and velocities,
%, the normal part of the elastic repulsive force is given as $F_{N}^{e,ij}=-k_{n}\delta n_{ij}$ where $k_n$ is the normal stiffness. This force acts when the overlap distance between two particles, $\delta n_{ij} < 0$. In addition to the elastic part, a normal viscous damping is also added to obtain the static equilibrium in reasonable time. Similarly for the tangential part, the force,$F_{T}^{ij}$ varies linearly with the tangential overlap until the initiation of sliding which takes place when $F_{T}^{ij}\geq\pm\mu F_{N}^{e,ij}$ where $\mu$ is the sliding friction coefficient. A constant attractive force of magnitude $F_0$ acting center to center between two particles is introduced when a contact is made. It should be noted that the Coulomb inequality is applicable to the repulsive elastic normal force only. The rolling resistance between the particles is incorporated\cite{pantina2005elasticity,furst2007yielding}.
%The total torque,$M_{ij}$ on a particle is calculated by summing up the torque exerted by the tangential contact force and the opposing rolling torque, $M_{ij}^{r}$,
%\begin{eqnarray}
%M_{ij}=-\frac{F_{T}^{ij}D}{2} + M_{ij}^{r}.
%\end{eqnarray}
%The opposing rolling torque is linearly proportional to the angular displacement of a torsional spring. The particle will roll if the total torque exceeds a threshold torque.
\begin{eqnarray}
m_i \frac{d^2 \B x_i}{dt^2} =\B F_i -m_i\gamma_t \frac{d\B x_i}{dt} +\B f_i(t) ;\\
I_i \frac{d^2 \B \theta_i}{dt^2} =\B T_i -I_i\gamma_r \frac{d\B \theta_i}{dt} +\B L_i(t) \ 
\label{eq:lang}
\end{eqnarray}
where $\B f_i(t)$ and $\B L_i(t)$ are respectively a $\delta$-correlated random force/torque having zero mean with the following properties:
\begin{equation}
\langle \B f_i(t) \cdot \B f_j(t+\tau)\rangle = 2 \Gamma \delta (\tau) \delta_{ij};\\\
\langle \B L_i(t) \cdot \B L_j(t+\tau)\rangle = 2 \Gamma \delta (\tau) \delta_{ij}
\end{equation}
Here $\gamma_t$ and $\gamma_r $ are translational and rotational damping coefficient, $m_i$ and $I_i$ are the mass and moment of inertia of the particle $i$, $\B F_i$ and $\B T_i$  are the force and torque due to inter-particle interactions, $\Gamma$ is the strength of fluctuations and $\B x_i$ and $\B \theta_i$  denote the translational and rotational degrees of freedom.

\begin{figure}[ht!]
\begin{center}
\includegraphics[scale=0.11]{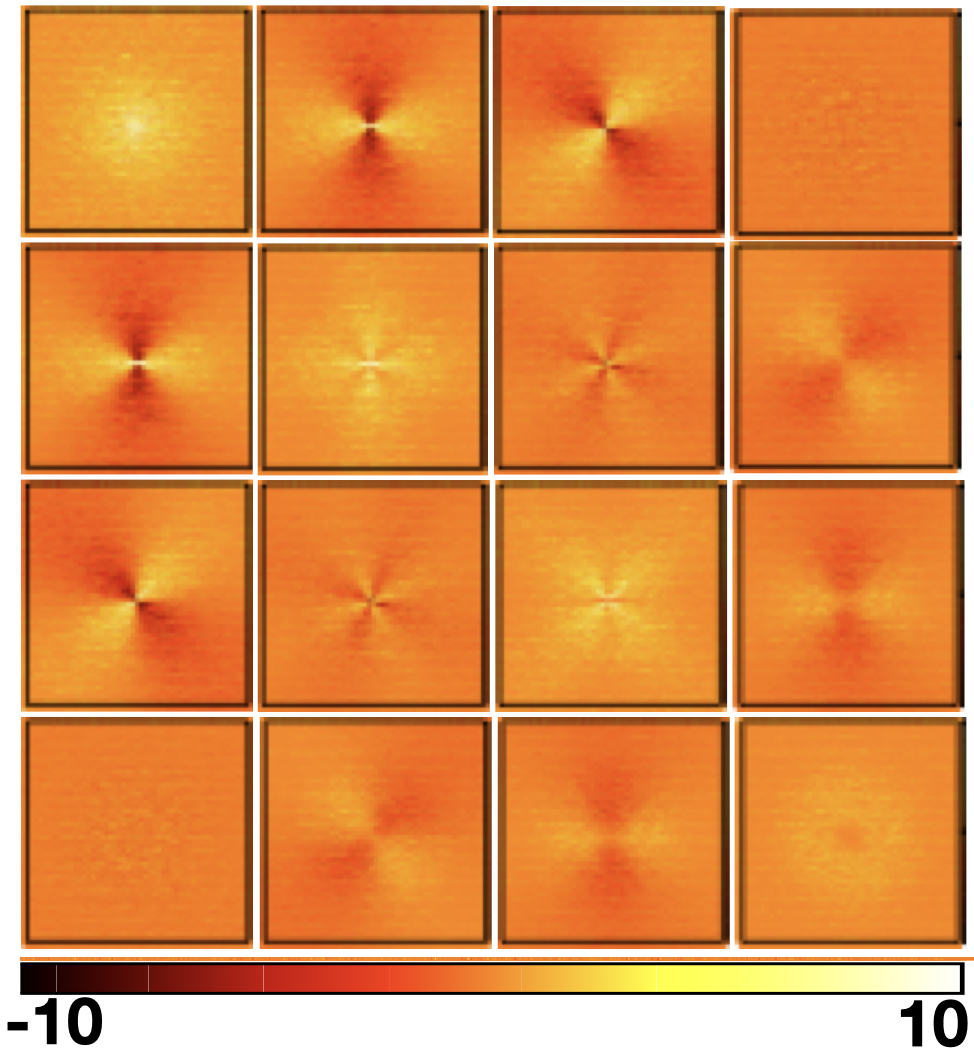}
\includegraphics[scale=0.11]{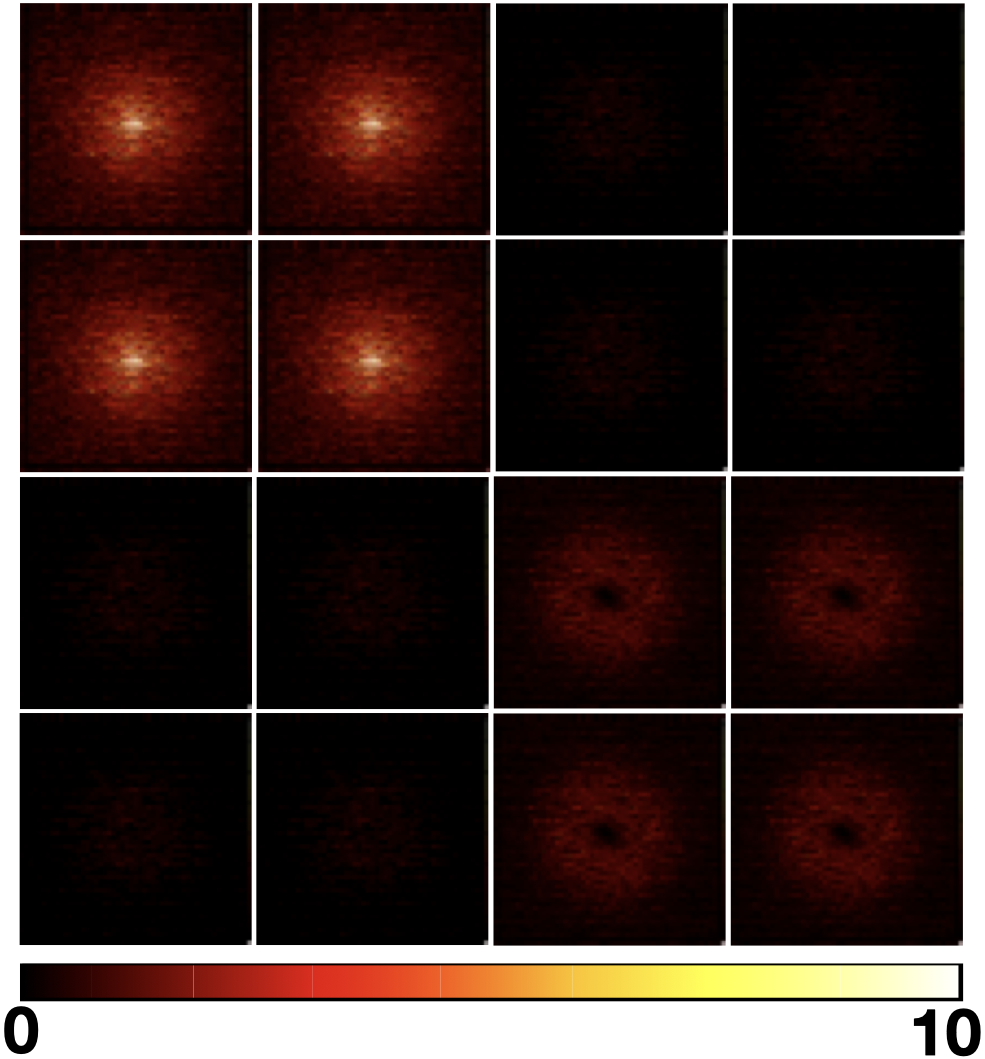}
\caption{Left:Fourier space stress correlation fields in the Cartesian frame for $P=2.75$; Right: The same field in the Radial frame. }
%:$\uttilde{\widehat{C}}_{\vec q}$ fields at $P=2.75$
%Bottom Left:
%$\uttilde{\widehat{C}}_{\vec q}$ fields at $P=131.7$; Bottom Right:
%$\uttilde{\mathring{\widehat{C}}}_{\vec q}$ at $P=131.7$
\label{fig:lowroll}
\end{center}
\end{figure}

Thermal fluctuations will lead to the flocculation of particles as the interaction is attractive. 
We compress the gel network isotropically and quasistatically at very slow strain rate to various target pressures. The system is then allowed to reach mechanical equilibrium(i.e total force and torque on each particle vanish, $O(10^{-7})$ with almost zero kinetic energy, $O(10^{-16})$) at different target pressures before we measure the stress correlations. As the compression proceeds, the flocs will connect to each other and form a percolating network at gel point. 
%The formation of a system spanning network is signaled by positive macroscopic  pressure. The macroscopic stress tensor,$\sigma_{\gamma \delta}$ is given by,
%\begin{equation}
%\sigma_{\gamma \delta} =\frac{1}{V}\sum_{j\neq i}\frac{r^{\gamma}_{ij} F^{\delta}_{ij} }{2}
%\end{equation}
%where $r^{\gamma}_{ij}$ and $F^{\delta}_{ij}$ are respectively $\gamma$ and $\delta$ Cartesian components of branch vector and contact force. The pressure $P$ will be given by the trace of the stress tensor. 
%The measurements are done at different pressures over ten independent configurations. 
From the particle level force data, we construct the stress field and calculate all the correlations between different stress components. 
%The coarse-grained stress tensor $\tensor\sigma(\vec r)$ of this system is defined as ~\cite{goldhirsch2002microscopic}:
%\begin{equation}\label{eq:gg}
%{\sigma}_{\gamma\delta}(\vec r)=-\frac{1}{2}\sum_{i,j; i\ne j} F_{ij}^\gamma r_{ij}^\delta\int_0^1\d s\,\phi(\vec r-\vec r_i+s\vec r_{ij})
%\end{equation}
%where $\gamma$, $\delta$ denotes the Cartesians coordinates, $F_{ij}$ is the interparticle force, $r_i$ is the position of particle, $i$, $r_{ij}$ is the radial vector between particles $i$ and $j$ and $\phi$ is some coarse-graining function that integrates (in two dimension) to unity and vanishes after a cut-off length, $r_c$. When $\phi$ is taken as 2D delta function (Hardy's microscopic stress~,
The stress tensor in Fourier space\cite{evans2008statistical}, reads:
\begin{equation}\label{eq:hardy}
\sigma_{\gamma\delta {\vec q}}=\frac{1}{2A}\,\sum_{i,j,i\ne j}\,F_{ij}^{\gamma}r_{ij}^{\delta}\frac{e^{-i{\vec q}\cdot {{\vec r}_i}}-e^{-i{\vec q}\cdot {{\vec r}_j}}}{i\vec q\cdot {\vec r}_{ij}}
\end{equation}
where $A$ is the system area, $\vec q$ is allowed wave vector,$\gamma$, $\delta$ denotes the Cartesian coordinates, $\vec F_{ij}$ is the inter-particle force, $\vec r_i$ is the position of particle $i$, $\vec r_{ij}$ is the radial vector between particles $i$ and $j$.
As the stress tensor is non-symmetric because of non-central interactions, this representation will have four spherical components in Fourier space\cite{lemaitre2021stress}:$\sigma_{1 \vec q}=-\frac{1}{2}\left(\sigma_{xx\vec q}+\sigma_{yy\vec q}\right)$;
$\sigma_{2 \vec q}=\frac{1}{2}\left(\sigma_{xx\vec q}-\sigma_{yy\vec q}\right)$;
$\sigma_{3 \vec q}=\frac{1}{2}\left(\sigma_{xy\vec q}+\sigma_{yx\vec q}\right)$;
$\sigma_{4 \vec q}=\frac{1}{2}\left(\sigma_{xy\vec q}-\sigma_{yx\vec q}\right).$\\
%\begin{figure}[h!]
%\begin{center}
%\includegraphics[scale=0.14]{AnaelFig2}
%\includegraphics[scale=0.14]{AnaelFig3}\\
%\includegraphics[scale=0.1]{AnaelFig5a}
%\includegraphics[scale=0.1]{AnaelFig5b}
%
%\caption{Top Left:The real-valued fields ${{\widehat{C}}}_{\vec q\,ab}$ for granular system are displayed as a (symmetric) matrix. Top Right: real part of the different components of the radial spherical autocorrelation matrix.
%Bottom left: Pressure autocorrelation function vs $q$ for $\mu=1$(squares and circles representing two different cases where in one case Coulomb friction cut off is reached abruptly and in another it is reached smoothly. Triangles represent frictionless case. Bottom right:Torque autocorrelation function vs $q$ for $\mu=1$}
%\label{fig:C}
%\end{center}
%\end{figure}
\textit{Stress correlations and isotropy}: If the system is translation-invariant, the autocorrelation matrix of these Cartesian spherical (CS) stress components in Fourier space reads as $\uttilde{S}_{\vec q}=\frac{1}{A}\,\left\langle\utilde{\sigma}_{\vec q}\,\utilde{\sigma}_{\vec q}^*\right\rangle_c$,
%\begin{equation}
%\label{eq:fourier:cartesian:corr}
%\uttilde{S}_{\vec q}=\frac{1}{A}\,\left\langle\utilde{\sigma}_{\vec q}\,\utilde{\sigma}_{\vec q}^*\right\rangle_c
%\end{equation}
where ${}^*$ denotes the complex conjugate and $\left\langle\ \right\rangle_c$ represents the second cumulant for the ensemble average. $\uttilde{S}_{\vec q}$ is calculated for all the stress components defined earlier.
 We show the same in Fig.~\ref{fig:lowroll} (Left) as a matrix of fields for pressure, $P=2.75$ which is very close to the gel point. We observe similar fields for $P=328$ (not shown). Only pressure, $S_{11\,\vec q}$ and torque, $S_{44\,\vec q}$ auto correlations and their cross correlations are isotropic and the rest are anisotropic. The whole matrix is clearly symmetric with unexpected anisotropic cross correlations between the torque and other fields as previously observed in granular matter\cite{lemaitre2021frictional,lemaitre2021stress}. Although the gel network is formed at very low packing fraction with fractal correlations, it is very surprising  that the structure of the stress auto-correlation fields bears a striking resemblance to that in frictional granular matter. 
To understand the role of material isotropy, we now calculate the Fourier space auto-correlations matrix in radial frame with basis ($\vec e_q,\vec e_\phi$). Corresponding stress vector components are given as: $\sigma^{\uvec q}_{1\,\vec q}=-\frac{1}{2}\left( \sigma_{qq\,\vec q}+ \sigma_{\phi\phi\,\vec q}\right)$;
$\sigma^{\uvec q}_{2\,\vec q}=\frac{1}{2}\left( \sigma_{qq\,\vec q}- \sigma_{\phi\phi\,\vec q}\right)$;
$\sigma^{\uvec q}_{3\,\vec q}=\frac{1}{2}\,\left(\sigma_{q\phi\,\vec q}+\sigma_{\phi q\,\vec q}\right)$;
$\sigma^{\uvec q}_{4\,\vec q}=\frac{1}{2}\,\left(\sigma_{q\phi\,\vec q}-\sigma_{\phi q\,\vec q}\right)$ where $\uvec q\equiv \vec q/q$ denotes the direction vector in reciprocal space. 
%As earlier, these radial components define a vector,
%$\utilde{\widehat{\sigma}}^{\uvec q}_{\vec q}=(\widehat{\sigma}_{1\,\vec q}^{\uvec q},\widehat{\sigma}_{2\,\vec q}^{\uvec q},\widehat{\sigma}_{3\,\vec q}^{\uvec q},\widehat{\sigma}_{4\,\vec q}^{\uvec q})$.
The autocorrelation matrix of these radial components is given by:
$\uttilde{\mathbb{S}}_{\vec q}=\frac{1}{A}\langle \utilde{\sigma}_{{\vec q}}^{\uvec q}(\utilde{\sigma}_{{\vec q}}^{\uvec q})^*\rangle_c$. It can be easily shown \cite{lemaitre2021stress} that under the constraint of material balance and isotropy, the structure of the radial autocorrelation fields contain just two spatially isotropic functions, namely pressure and torque auto-correlations. 
%The corresponding form is:
%\begin{equation}
%  \label{axial3}
%\uttilde{\mathring{\widehat{C}}}=
%\left(
%  \begin{matrix}
%    \mathring{\widehat{C}}&\mathring{\widehat{C}}&0&0\\
%    \mathring{\widehat{C}}&\mathring{\widehat{C}}&0&0\\
%    0&0&\mathring{\widehat{C}}'&\mathring{\widehat{C}}'\\
%    0&0&\mathring{\widehat{C}}'&\mathring{\widehat{C}}'\\
%  \end{matrix}
%\right)
%\end{equation}
%%%%%%%%%%%%%%%%%%%%%%%%%%%%%%%%%%%%%%%%%%%%%%%%%%%%%%
\begin{figure}
		 \includegraphics[scale=0.15]{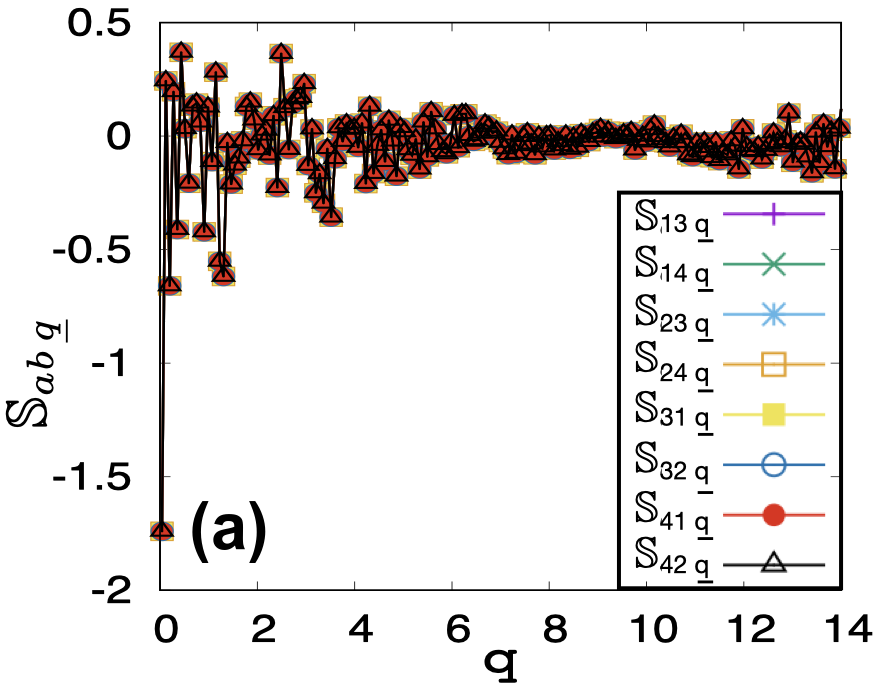}
		 \includegraphics[scale=0.15]{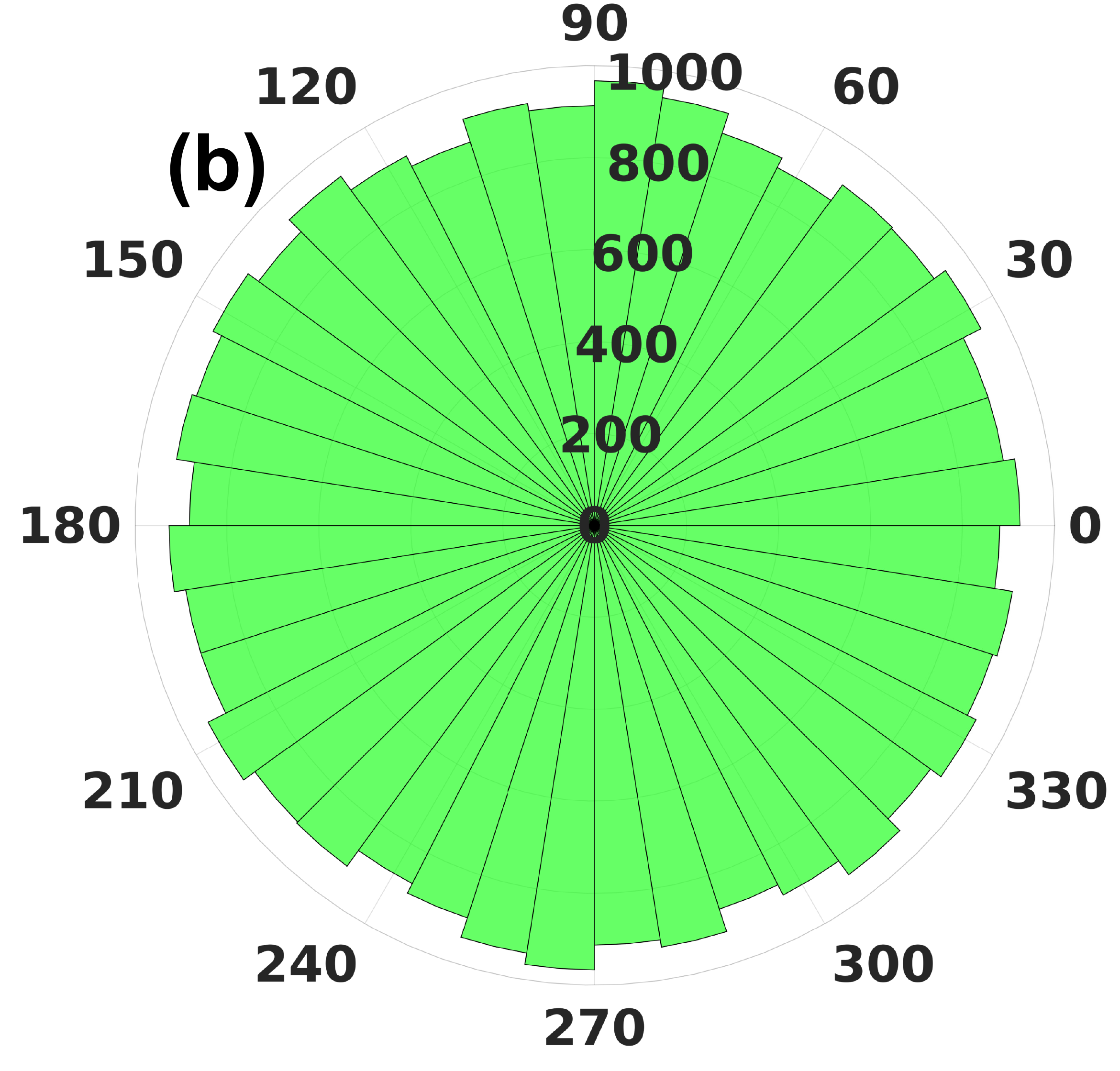}
  \caption{(a) Angle averaged off-diagonal correlations ${\mathbb{S}}_{ab\, \vec{q}}$, $ab=13,14,23,24,31,32,41,42$ for $P=2.75$ (b)Contact orientation distribution at $P=2.75$. Same distribution is observed at $P=328$.(not shown)}
	\label{offdiaAnis}
	\end{figure}
%%%%%%%%%%%%%%%%%%%%%%%%%%%%%%%%%%%%%%%%%%%%%%%%%%%%%%%%%%%% 
%where $\mathring{\widehat{C}}(q)$ and $\mathring{\widehat{C}}'(q)$ are the pressure and torque autocorrelations. 
The same field is shown in  Fig.~\ref{fig:lowroll} (Right) depicting excellent agreement with the theory which implies that the constraints of mechanical balance and material isotropy are indeed satisfied in our system. The off-diagonal components, which are expected to be identically zero from theory, show remnant fluctuations. In previous studies\cite{lemaitre2021stress}, these residual fluctuations were attributed to numerical inaccuracy which is indeed true in the present study as angle-averaging (see Fig.~\ref{offdiaAnis} (a)) strongly reduces the fluctuations revealing their random nature. We also compute the contact orientation distribution and find that the contact network remains isotropic throughout the deformation (see Fig.~\ref{offdiaAnis} (b)). This provides additional proof of material isotropy.

%\begin{figure}[h!]
%\begin{center}
%\includegraphics[scale=0.14]{lowP_fabric.pdf}
%\includegraphics[scale=0.14]{highP_fabric.pdf}
%\caption{Left:Contact orientation distribution at $P=2.75$; Right:Contact orientation distribution at $P=328$.}
%\label{fig:Anis}
%\end{center}
%\end{figure}

\begin{figure}[h!]
\includegraphics[scale=0.15]{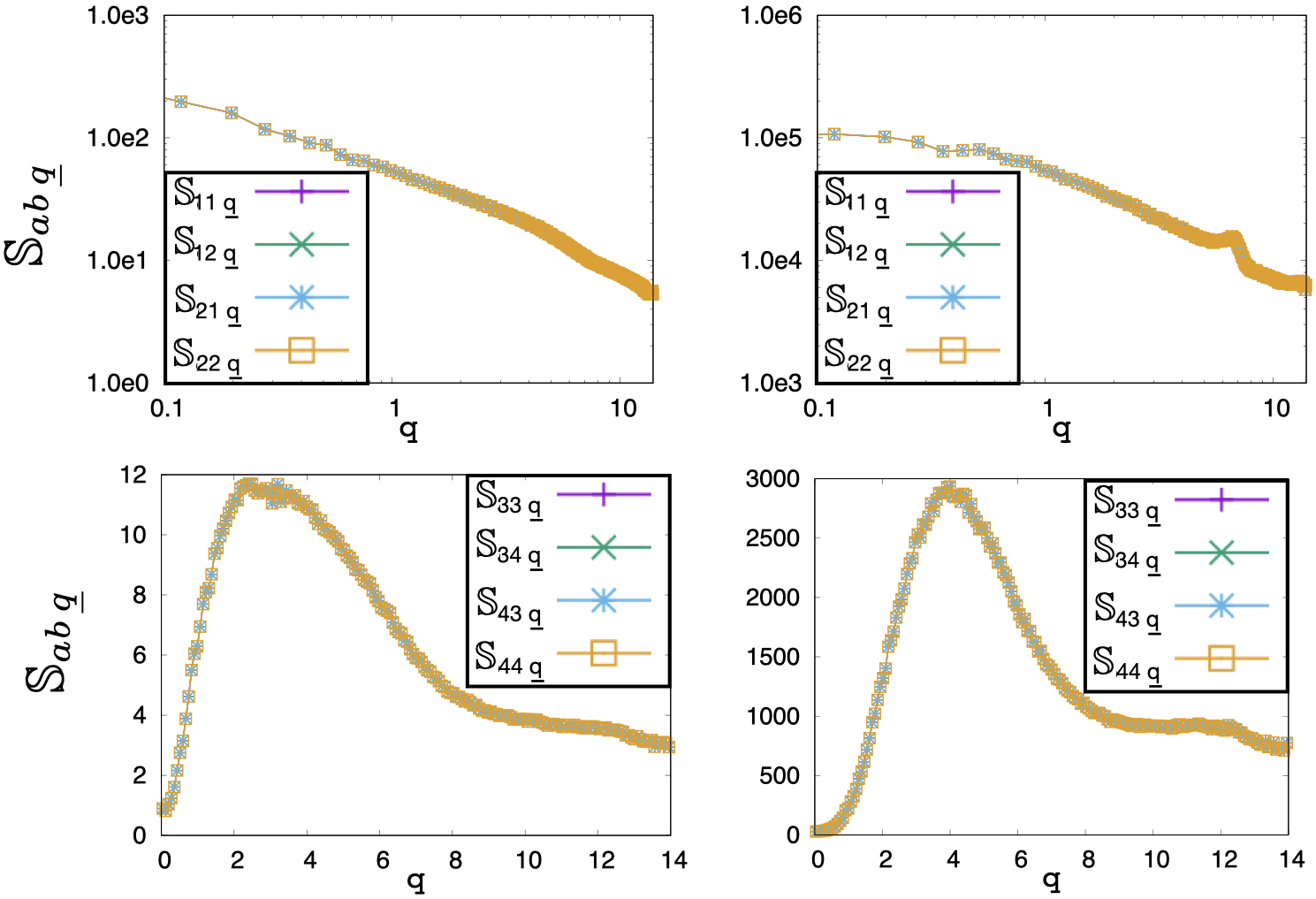}
\caption{Top: plot of the pressure autocorrelation function ${\mathbb{S}}_{ab\, \vec{q}}$, $ab=11, 12, 21, 22$ for $P=2.75$ (Left) and $P=328$ (Right). 
Bottom: plot of the torque density autocorrelation function ${\mathbb{S}}_{ab\, \vec{q}}$, $ab=33, 34, 43, 44$ for $P=2.75$ (Left) and $P=328$ (Right)}
\label{fig:pretor}
\end{figure}
Next, we investigate the pressure and torque auto-correlations (Fig.~\ref{fig:pretor}).  The torque correlations approaches zero in the $q\to 0$ limit for high pressure whereas it reaches a finite non-zero value for the low pressure. It implies that the gel network carries some long range torque correlations close to the gel point and far away from it, the torque fluctuations become hyperuniform. The pressure correlations show divergence :($q^{-\nu}$) in the small wave-number limit for low pressure whereas it approaches almost a constant value as $q\to0$ for high pressure. The divergence exponent is measured by averaging over all the components of relevant correlations($11,12,21,22$) and it turns out that the pressure auto-correlations show a divergence of the form $q^{-0.51}$ at $P=2.75$. \\
\textit{Pressure fluctuations}: In earlier works\cite{lemaitre2021stress}, it was shown that the normal elastic behaviour demands normal fluctuations of pressure and torque density. 
%For the case of frictional granular matter, the torque density fluctuations are hyperuniform but the pressure fluctuations decay is slower than normal leading to divergence in the pressure correlations as $q\to0$. 
The torque auto-correlations are not hyperuniform for the colloidal gel at low pressure. However, at high pressure $P=328$ , torque auto-correlations become hyper-uniform and thus the stress autocorrelation is fully determined by the pressure fluctuations statistics. So we compute the pressure fluctuations in real space by placing randomly a circular window of radius $R$ and measuring the variance of pressure, $\Delta_P(R)$ due to circle-to-circle and sample to sample fluctuations,
\begin{equation}
\Delta_P(R) \equiv \langle P(R)^2 \rangle - \langle P(R)\rangle ^2 \sim \frac{1}{R^\eta}\ .
\label{eqvar}
\end{equation}
In case of normal fluctuations, the variance is expected to decay as the inverse area of the probing window i.e $1/R^2$ and consequently the stress or pressure correlation shows no divergence in the low wave number limit. 
\begin{figure}[h!]
\includegraphics[scale=0.13]{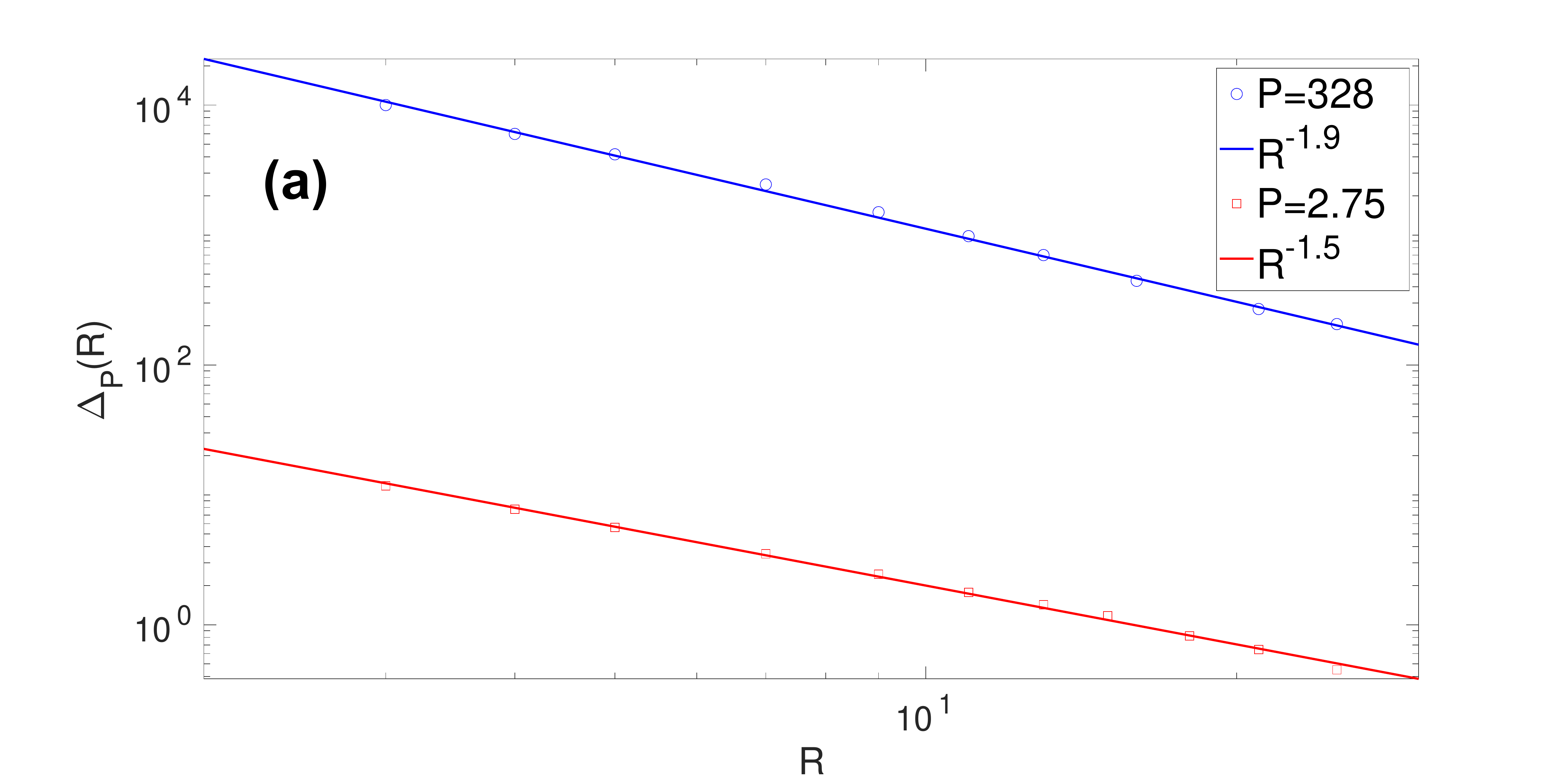}
\includegraphics[scale=0.14]{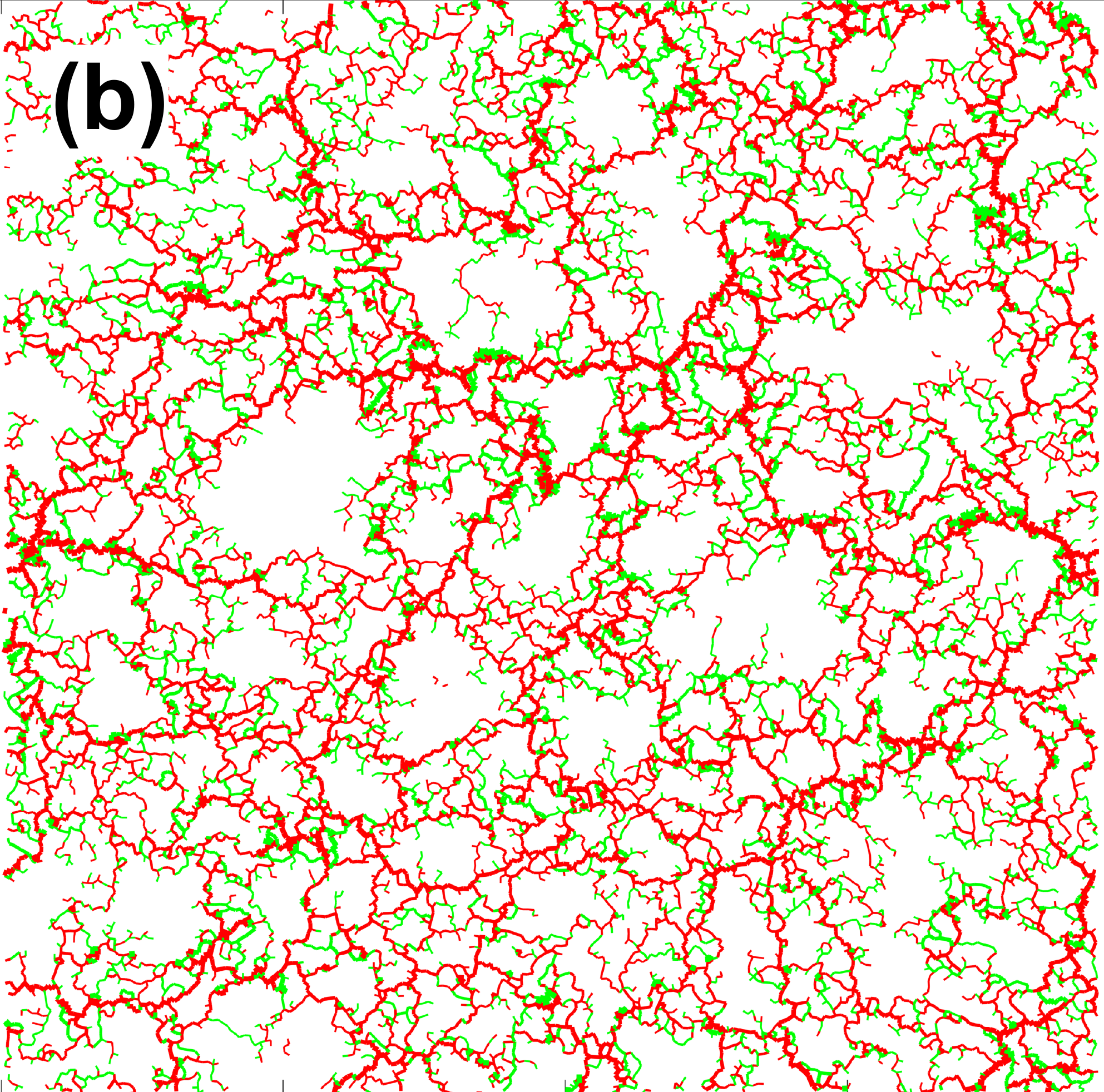}
\includegraphics[scale=0.14]{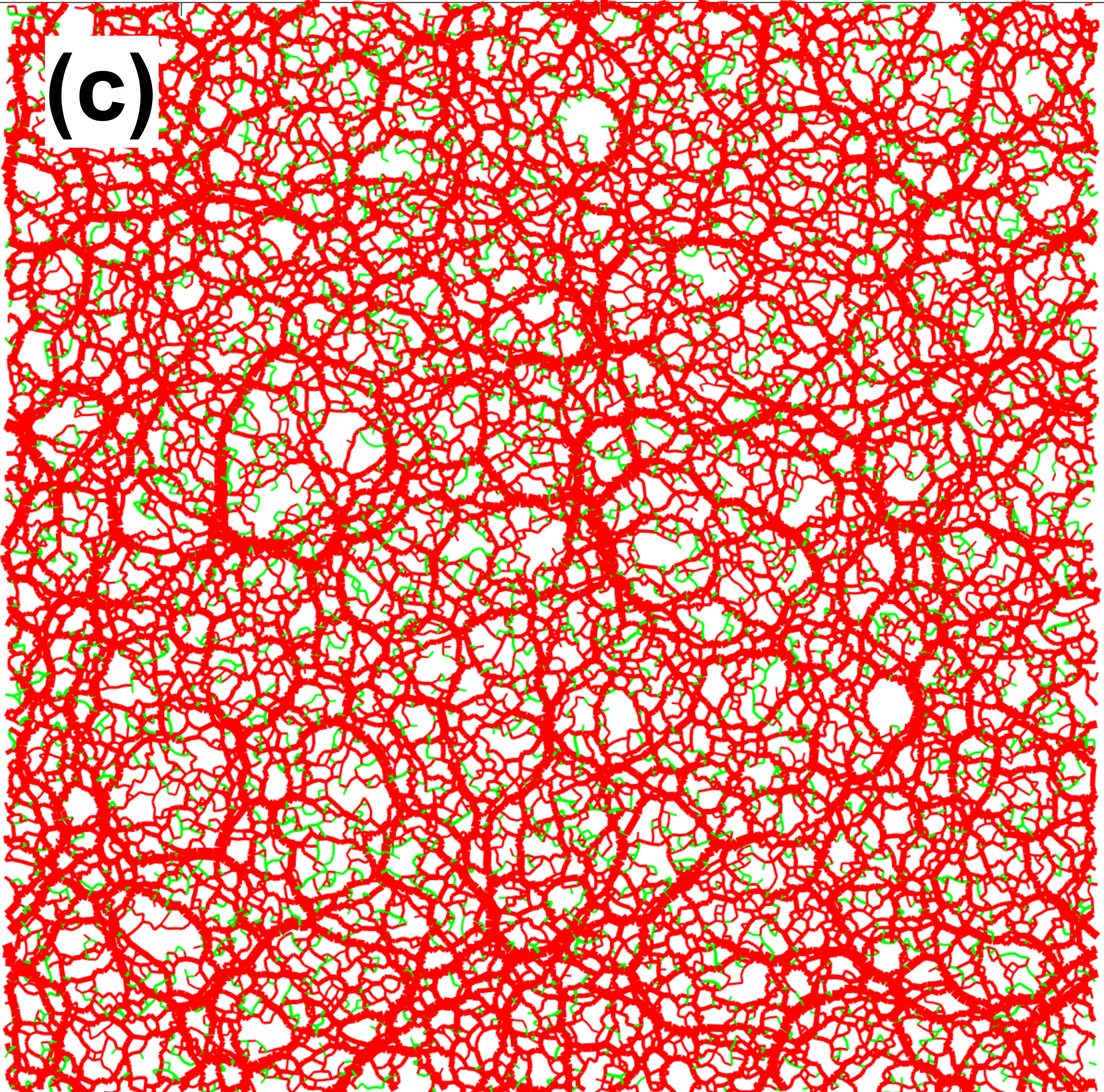}
\caption{(a)Pressure variance as a function of radius of the probing circular window. The slope of the straight lines are $-1.5$ and $-1.9$ respectively for low and high pressure.(b)Force chains in colloidal gel sample, $P=2.75$, (c) $P=328$. The compressive force is drawn in red color whereas the green color represents the tensile force.The line thickness is scaled as per the magnitude of the pairwise force.}
\label{variance}
\end{figure}
%%%%%%%%%%%%%%%%%%%%%%%%%%%%%%%%%%
In more generic sense, the exponent of the divergence in the Fourier space is related to the exponent of the decay of the pressure fluctuations in real space as $\nu=2-\eta$.\cite{lemaitre2021stress} In Fig.~\ref{variance} (a) we show the pressure fluctuations as a function of $R$ for both pressures, $P=2.75$ and $P=328$. For low pressure, $\Delta_P(R)$ decays like $R^{-1.5}$ which is slower than the normal decay. This slope is in line with the slope of the divergence in the stress correlations in Fourier space as per Eqn.$\left(5\right)$. Note that although the torque correlations are not hyperuniform at low pressure, the long range stress correlations are predominantly decided by the pressure fluctuations. Far above the gel point, the pressure fluctuations become approximately normal with a slope of $1.9$ leading to almost divergence free behaviour of the pressure auto-correlations in the small $q$ limit. Again it is the statistics of the pressure fluctuations which determine the long distance decay of the stress correlations. \\
\textit{Force chains}: In order to understand the microscopic origin of the long distance decay behaviour for low and high pressures, we plotted the real space map of the force chains in Fig.~\ref{variance}(b)-(c). One can clearly observe the inhomogenity in the force chain network at low pressure created due to the presence of large voids in the gel contact network. Note that the gel network is expected to be fractal in nature at low pressure and the inhomogenity in the force chain network is created due to the strong fractal like structural correlations whereas the frictional granular media is not fractal but yet gives rise to strong heterogeneity in the force chains. As the frictional granular matter jams at high packing fractions, these anomalous correlations once formed do not relax through structural rearrangement. In contrast, under compression the gel network looses fractal correlations and undergoes structural rearrangement through collapse of voids leading to homogeneous and isotropic force chain network at high pressures. 
Following our previous works\cite{roy2016yielding}, we also determine the fractal properties of the network. Our data suggests a fractal dimension of $1.44$. It is important to note that this fractal dimension is some average dimension over the packing fractions, not the true dimension at each packing fraction. Recent simulation \cite{seto2013compressive} reveals that the fractal dimension changes with the packing fraction and the network is no longer fractal at high pressure.\\
\begin{figure}[ht!]
\includegraphics[scale=0.15]{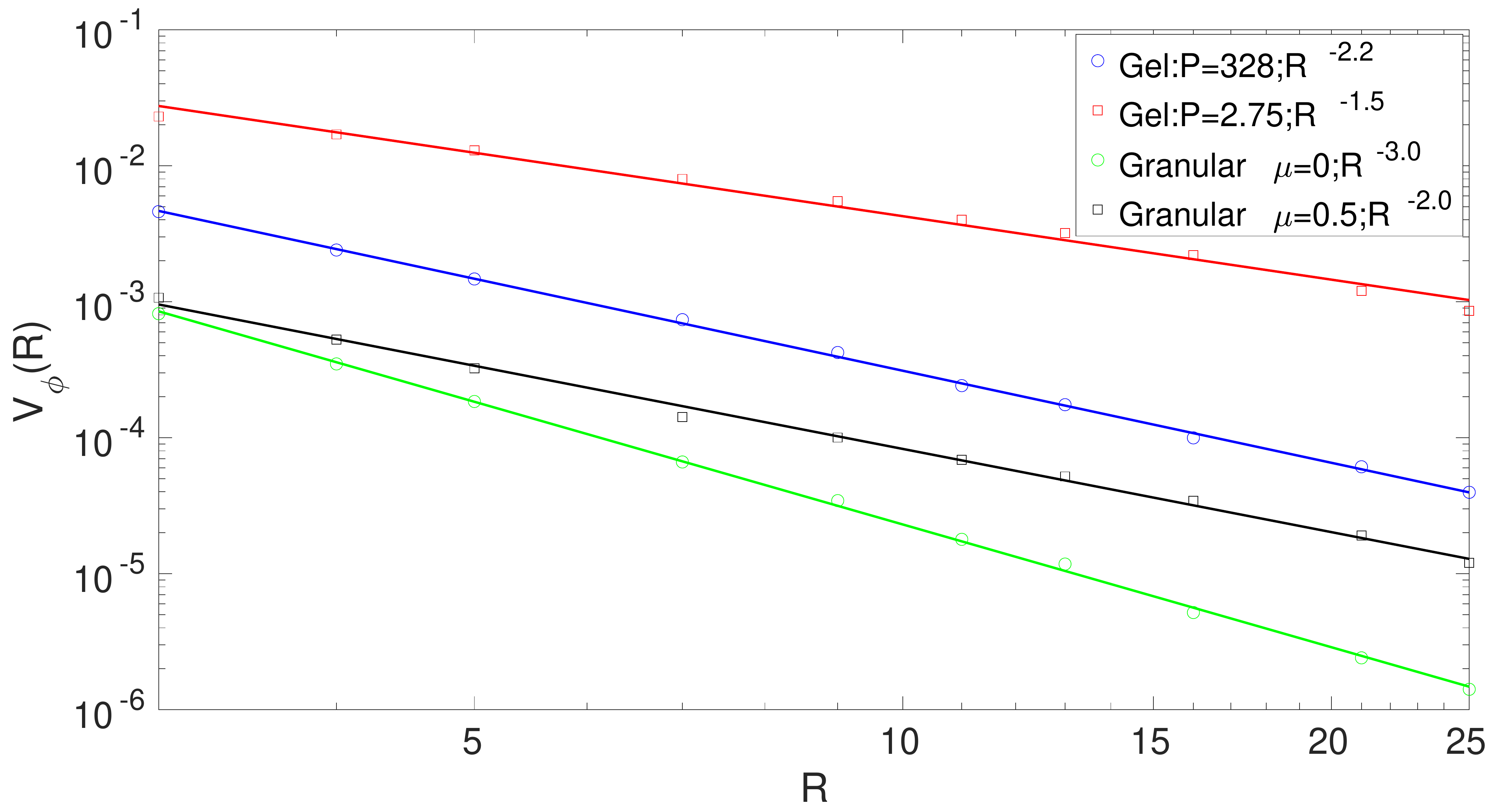}
\caption{Variance of the local packing fraction, $V_{\phi}$ vs $R$ as observed in granular materials and gels.}
\label{fig:hyperuniform}
\end{figure}
%%%%%%%%%%%%%%%%%%%%%%%%%%%%%%%%%%
\textit{Local volume fraction fluctuations and hyperuniformity}: In order to understand the microscopic origin of the force heterogeneities under a common framework, we compute the local volume fraction fluctuations as it provides the appropriate structural descriptions of these disordered packings.\cite{zachary2011hyperuniformity,torquato2018hyperuniform} We place a circle of radius $R$ randomly within the system and measure the local volume fraction within this window. Similar to the pressure variance, we measure the variance of the packing fraction, $V_{\phi}(R) \equiv \langle \phi(R)^2 \rangle - \langle \phi(R)\rangle ^2 \sim \frac{1}{R^{d+\alpha}}$
%\begin{equation}
%V_{\phi}(R) \equiv \langle \phi(R)^2 \rangle - \langle \phi(R)\rangle ^2 \sim %\label{eqvar2}
%\end{equation}
where exponent $\alpha$ expresses the degree of uniformity.\cite{torquato2018hyperuniform}For normal uniformity, $\alpha=0$, for hyperuniform systems (disordered isotropic systems where long wavelength fluctuations are suppressed) $\alpha>0$ and $\alpha<0$ denotes hyper-fluctuations observed in fractal networks. We also carried out simulation on frictional and frictionless granular materials \cite{lemaitre2021frictional,lemaitre2021stress} to measure the real space packing fraction fluctuations. In Fig.~\ref{fig:hyperuniform}, we plot the variance of the local packing fraction, $V_{\phi}$ as a function of $R$ for colloidal gel ($P=2.75$ and $P=328$; here $\mu=0.5$) and the granular materials ($\mu=0$ and $\mu=0.5$; here $P=72$). The colloidal gel network at low pressure shows hyper-fluctuations i.e the local packing fraction fluctuations decay slower than the reciprocal of the area of the observation window. The exponent of this anomalous decay is same as that of the decay of the pressure variance which suggests that the fractal like structural correlations control the pressure fluctuations close to the gel point. The gel network at high pressure shows hyperuniformity as the exponent of the decay of $V_{\phi}$ is greater than $2$. Similarly, the frictionless granular materials also show hyperuniform behaviour with respect to the local packing fraction fluctuations ($R^{-3}$ decay). Lastly the frictional granular material shows {\it  uniform behaviour} like $1/R^2$ decay. This is the most interesting result of this work as it links the abnormal pressure fluctuations to the non-hyperuniform behaviour of the many particle systems. The normal elastic decay of stress correlations is intimately linked to the long wavelength suppression of the local packing fraction fluctuations i.e hyperuniform behaviour. \\
%This result also suggest the strong correlations between stress and volume fluctuations both of which need to be taken into account simultaneously while constructing the statistical ensemble i.e total entropy is not just the sum of configurational and stress entropies\cite{blumenfeld2012interdependence,bi2015statistical}. \\
\textit{Summary}: In conclusion, we convincingly demonstrate that under the mechanical balance and material isotropy constraints, the full stress autocorrelation matrix in soft gel materials is determined by the torque and pressure auto-correlations similar to frictional granular matter, thus suggesting an universal behaviour across diverse amorphous solids. Interestingly, close to the gel point, we observe the divergence in pressure correlations in the $q\to0$ limit and non-hyperuniform torque correlations whose contribution to the stress correlations at long length scale is subdominant and the pressure fluctuation statistics strongly determine the exponent of the divergence. We have also linked this divergence to inhomogenous and anisotropic force networks formed due to the presence of large voids at low packing fractions. 
%It is also shown, at this point the fractal like structural correlations robustly control the pressure fluctuations. 
At high pressure, the system shows almost divergence free elastic behaviour with hyperuniform torque fluctuations and the normal pressure decay as the large voids collapse and the system has enough space to relax the force inhomogenities which is not possible for the frictional granular matter jammed at high packing fractions. Most importantly, we identify a connection between the  abnormal pressure fluctuations and the non-hyperuniform behaviour of the system with respect to the local packing fraction fluctuations, thus establishing a common framework to understand the deviations from the normal elastic like behaviour across different non-equilibrium systems. The result of this work begs for the development of a novel theoretical framework to understand the mechanical response of gel and granular materials as these systems clearly deviate from the normal elastic behaviour as shown.\\
\textit{Acknowledgement-}S.R. acknowledges the support of SERB under Grant No. SRG/2020/001943. CM acknowledges SERB for financial support.

\bibliography{ref2}

\end{document}